\documentclass[aps,prl,twocolumn,showpacs]{revtex4-1}
\usepackage[dvips,pdftex]{graphicx}
\usepackage{hyperref}
\usepackage{amsmath}
\usepackage{color}

\begin{document}
\title{Localization in active incommensurate arrays}

\author{T.V.~Laptyeva$^1$, S.V.~Denisov$^{2,3}$, G.V.~Osipov $^1$ and M.V.~Ivanchenko$^3$}

\affiliation{$^1$ Theory of Control Department, Lobachevsky State University of Nizhny Novgorod, Russia \\
$^2$ Department of Theoretical Physics, University of Augsburg, Germany \\
$^3$ Department of Bioinformatics, Lobachevsky State University of Nizhny Novgorod, Russia}

\begin{abstract}
In a dissipationless linear lattice, spatial disorder or incommensurate modulation induce localization of the lattice eigenstates
and block spreading of wave packets. Additionally, incommensurate arrays allow for the metal-insulator transition at a finite modulation amplitude already in one dimension. The addition of nonlinearity to the lattice Hamiltonian causes interaction between the eigenstates, which results in a slow packet spreading. We go beyond the dissipationless limit and consider nonlinear quasi-periodic arrays that are subjected to the dissipative losses and energy pumping. We find that there is a finite excitation of oscillations threshold in both metallic and insulating regimes. Moreover, excitation in the metallic and weakly insulating regime displays features of the second order phase transition to global oscillations, in contrast to disordered arrays. The Anderson attractor regime is recovered only in the limit of strong localization. The identified transition, and the further onset of chaos and synchronization can be potentially realized with polariton condensates lattices and cavity-QED arrays.
\end{abstract}


\maketitle
Localization of eigenstates and halt of wave propagation in spatially modulated potentials remains in the focus of studies for more that fifty years since the seminal paper of Anderson \cite{Mirlin,fifty}. From the current perspective, it can be obtained in various spatial inhomogeneities: an external dc field leads to Wannier-Stark localization and Bloch oscillations \cite{BO}, random yields Anderson localization \cite{PWA58}, and quasiperiodic modulation evokes Aubry-Andre localization \cite{AA}.  In the last decade its generic nature has been experimentally demonstrated  with electromagnetic \cite{Optics}, acoustic \cite{acoustic}, and matter waves \cite{BEC}. 
Recent theoretical advances in the problem of localization in presence of nonlinearity and interactions \cite{ND} receive encouraging confirmation in experimental optics of ultracold atoms \cite{nonlinear2,lucioni}. 

A novel dimension in the field is now opened by the recent discovery of essentially dissipative exciton-polariton condensation in quantum wells  \cite{Kasprzak,Balili,epBEC,P_reviews,carusotto}. Such condensation centers can be arranged in various 1D and 2D structures by artificially created or inherent spatial inhomogeneities \cite{Balili, Lai}, or by rotating ring-shaped optical potentials and switching to the co-moving frame \cite{amico,berloff}. Each center would be an active open system, balancing between excitation by an incoherent pumping source and decay due to own light radiation. The centers would also interact with their neighbors producing both Josephson and dissipative coupling, the former typically prevailing. The resulting excitation and lasing dynamics can be appropriately described by Ginzburg-Landau type equations \cite{GLE} and studied in the framework of nonlinear dynamics, which has already revealed rich behavior and pattern formation in spatially homogeneous condensation centers structures \cite{GPE,P_reviews}. 

These advances have lead us to pose the problem of localization in \textit{active} arrays, where 
pumping and dissipation join the old players, nonlinearity and disorder \cite{we_arxiv}. We have demonstrated that Anderson attractors -- sparse oscillation patterns of Anderson modes -- emerge. The Anderson model, however, has a specific trait of all states being localized by an arbitrary small disorder in one-dimensional arrays and does not capture neither extended (metallic) inhomogeneous regime, not the transition to localized (insulating) regime as the disorder depth is varied. In this light the Aubry-Andre model with incommensurate (quasiperiodic) instead of disordered potential has gained particular attention \cite{AA,Modugno}. Its straightforward experimental realization with optical potentials \cite{Modugno} has made it an attractive object of studying localization of ultracold atomic condensates in spatially modulated potentials \cite{BEC,nonlinear2,lucioni}.   

In this paper we study localization in active incommensurate arrays by using the discrete complex Ginzburg - Landau equation (GLE) \cite{dGLE} as a model. We find that the increase of the pumping strength leads to the second order type phase transition to complex oscillating patterns, both in the metallic and insulating regime. The average amplitude density scales as $(\alpha-\alpha^*)^\gamma, \ 1/2<\gamma<1$, approaching the upper limit, characteristic of the disordered lattice, with increasing modulation depth. The density of effectively excited sites displays a diverging derivative about the transition, manifesting concurrent delocalization. Finally, we show that the onset of multi-mode oscillations leads to dynamical chaos, with chaotic cluster  synchronization in the metallic and weakly localized regimes.

\begin{figure}[t]
\begin{center}
\includegraphics[width=\columnwidth,keepaspectratio,clip]{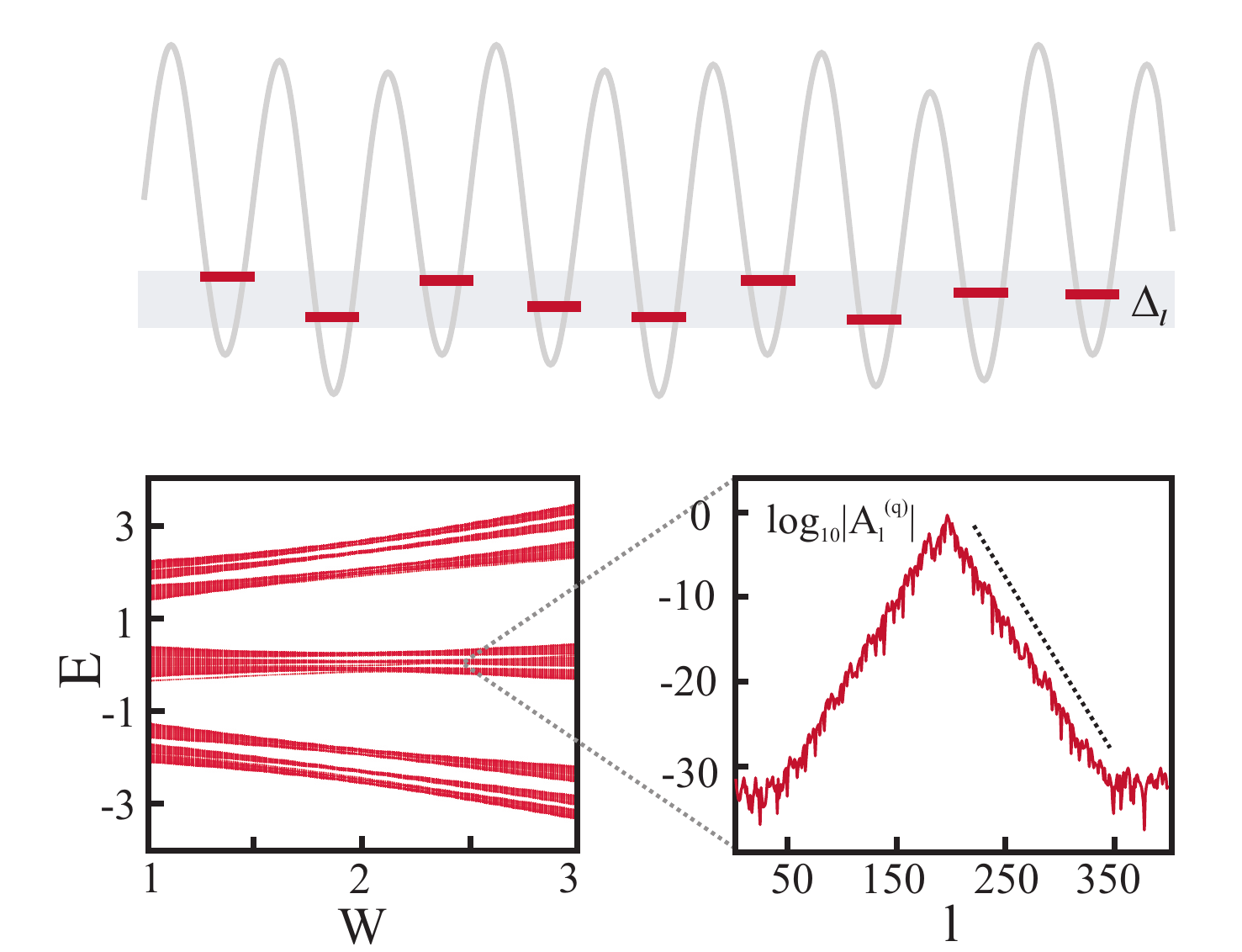}
\caption{(a) Schematic representation of the 1D Aubry-Andr\'e potential (\ref{eq:1c}). (b)
Eigenvalues $E = \lambda-2$ of the system (\ref{eq:1c}) obtained from numerical diagonalization as a function of the modulation strength $W$. (c) Typical exponentially localized eigenstate of (\ref{eq:1c}) in the insulating regime, $W=2.5$.} 
\label{fig:0}
\end{center}
\end{figure}

We consider a one-dimensional Ginsburg-Landau equation,
\begin{equation}
\label{eq:1a}
\begin{aligned}
&i\dot{z}_l=\Delta_l z_l + i\left(\alpha-\sigma\left|z_l\right|^2\right)z_l+\left|z_l\right|^2z_l\\
&-\left(1-i\eta\right)(z_{l+1}-2z_l+z_{l-1}),
\end{aligned}
\end{equation}
with incommensurate spatial inhomogeneity $\Delta_l = W\cos(2\pi\alpha l+\beta), \ \alpha=(\sqrt{5}-1)/2$, 
where $\beta\in[0, 2\pi]$ is some phase shift. Referring to the Anderson disordered model we will imply random uncorrelated and uniformly distributed $\Delta_l\in[-W/2,W/2]$. Further on, $\alpha$ is the pumping rate, $\sigma$ is the nonlinear dissipation coefficient, and $\eta$ is the strength of dissipative coupling between adjacent sites. 
Without loss of generality we set conservative nonlinearity and coupling coefficients to one. In numerics, periodic boundary conditions are assumed, $z_{N+1}=z_1$.

In the linear dissipationless limit, $\alpha=\eta=0$ and $\left|z_l\right|^2 \rightarrow 0$,
the stationary solutions $z_l=A_l {\rm e}^{ - i \lambda t}$  satisfy
\begin{equation}
\label{eq:1b}
\lambda_q A_l^{(q)}=\Delta_l A^{(q)}_l-A^{(q)}_{l+1}+2A^{(q)}_l-A^{(q)}_{l-1},
\end{equation}
which by $E \equiv \lambda-2$ reduces to the standard Aubry-Andre eigenvalue problem:
\begin{equation}
\label{eq:1c}
E_q A_l^{(q)}=\Delta_l A^{(q)}_l-A^{(q)}_{l+1}-A^{(q)}_{l-1},
\end{equation}
All eigenstates $A_{q,l}$  are extended for $W<2$ (which is called metallic regime) and exponentially localized for $W>2$ (insulating state), $|A^{(q)}_l|\sim \exp\left[-|l-l_q|/\xi \right]$, with the same localization length $\xi=[\ln(W/2)]^{-1}$, $l_q$,  $l_\lambda$ denoting the center of mass (Fig.\ref{fig:0}). 
\begin{figure}[t]
\begin{center}
\includegraphics[width=\columnwidth,keepaspectratio,clip]{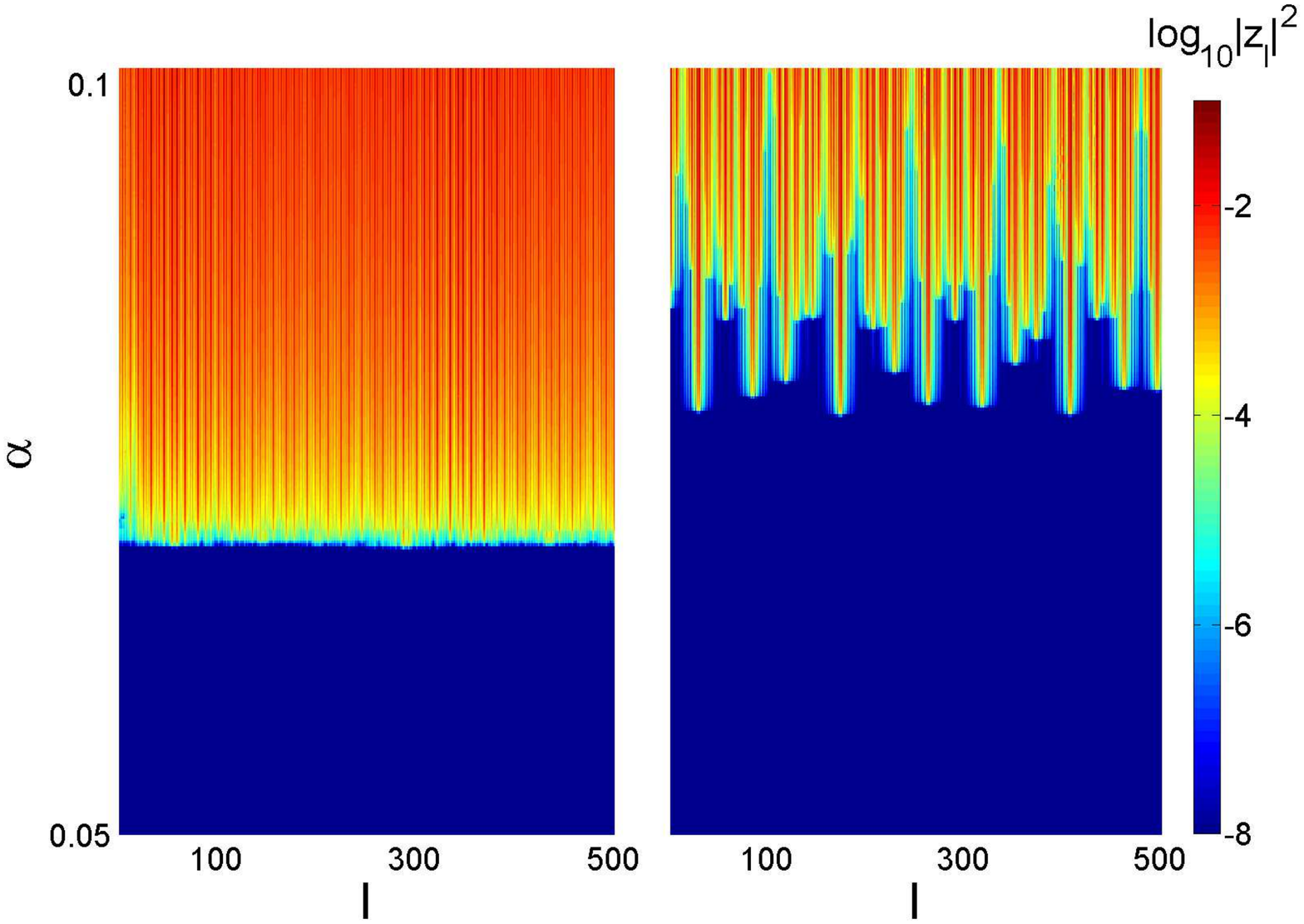}
\caption{(Color online) Excitation of oscillations in the incommensurate system (\ref{eq:1a}) upon the increase of the pumping for weak ($W=2.02$, left panel) and strong localization ($W=2.5$, right panel).  The color codes oscillation amplitudes at lattice sites, $\log_{10}|z_l|^2$, as functions of $\alpha$. The other parameters are $\eta=0.1, \ N=1000$.}
\label{fig:1}
\end{center}
\end{figure}

Considering evolution of the the net norm $Z=\sum |z_l|^2$ under Eq. (\ref{eq:1a}) one obtains \cite{we_arxiv}
\begin{equation}
\label{eq:4}
\dot{Z}=2\sum\left[(\alpha-\sigma|z_l|^2)|z_l|^2-\eta|z_{l+1}-z_l|^2\right].
\end{equation}
It follows that the zero solution $z_l \equiv 0$ is globally stable for all $\alpha \le 0$.
It also suggests that homogeneous in-phase solutions  $z_{l+1}\approx z_l$ 
are more energetically favorable than anti-phase ones, $z_{l+1}\approx -z_l$.


The linear stability of the zero solution against single-mode perturbations, $z_l(t)=\zeta A_l^{(q)}\exp[(p_q-i\lambda_q) t], \zeta\ll1$ is determined by the respective increments $p_q$. They can be calculated from 
Eq. (\ref{eq:1a}) and read
\begin{equation}
\label{eq:5}
p_q=\alpha-\eta\sum\left|{A}^{(q)}_{l+1}-{A}_l^{(q)}\right|^2.
\end{equation}
The  necessary stability condition of the zero solution is then $\mbox{max} \ p_q <0$. 
Note that this quantity  depends only on the modulation depth $W$, lattice phase $\beta$, 
and the ratio between incoherent pumping rate and dissipative coupling, $\alpha/\eta$. 
Moreover, it can be shown that the scaled excitation threshold 
\begin{equation}
\label{eq:6}
\frac{\alpha^*}{\eta}=\min\limits_{q}\frac{\alpha^*_q}{\eta}=\min\limits_{q}\sum\left|{A}^{(q)}_{l+1}-{A}_l^{(q)}\right|^2
\end{equation} 
is bounded, $0\le\alpha^*/\eta\le4$. 

Further, in the vanishing modulation limit $W\rightarrow 0$, one recovers the harmonic eigenstates $A_l^{q}\sim N^{-1/2}\exp(i\kappa l)$ and spectrum $\lambda_q=4\sin^2(\kappa/2), \ \kappa=\pi q/(2(N+1)), \ q=\overline{1, N}$. In this case the instability threshold (\ref{eq:6}) is minimized by the lower boundary mode $\lambda_q\rightarrow0, \ \alpha^*/\eta\rightarrow 0$ as $N\rightarrow\infty$. In the opposite limit $W\gg1$, deep in the insulating regime, the modes are essentially single-site localized and it follows $\alpha^*/\eta\sim1$. Hence, one can expect that in the metallic regime excitation of some mode (predictably, about the lower spectrum boundary) will produce extended oscillation patterns, while in the strongly insulating regime one may observe a complex pattern of localized oscillating spots, similar to Anderson attractors \cite{we_arxiv}.    

\begin{figure}[t]
\begin{center}
\includegraphics[width=\columnwidth,keepaspectratio,clip]{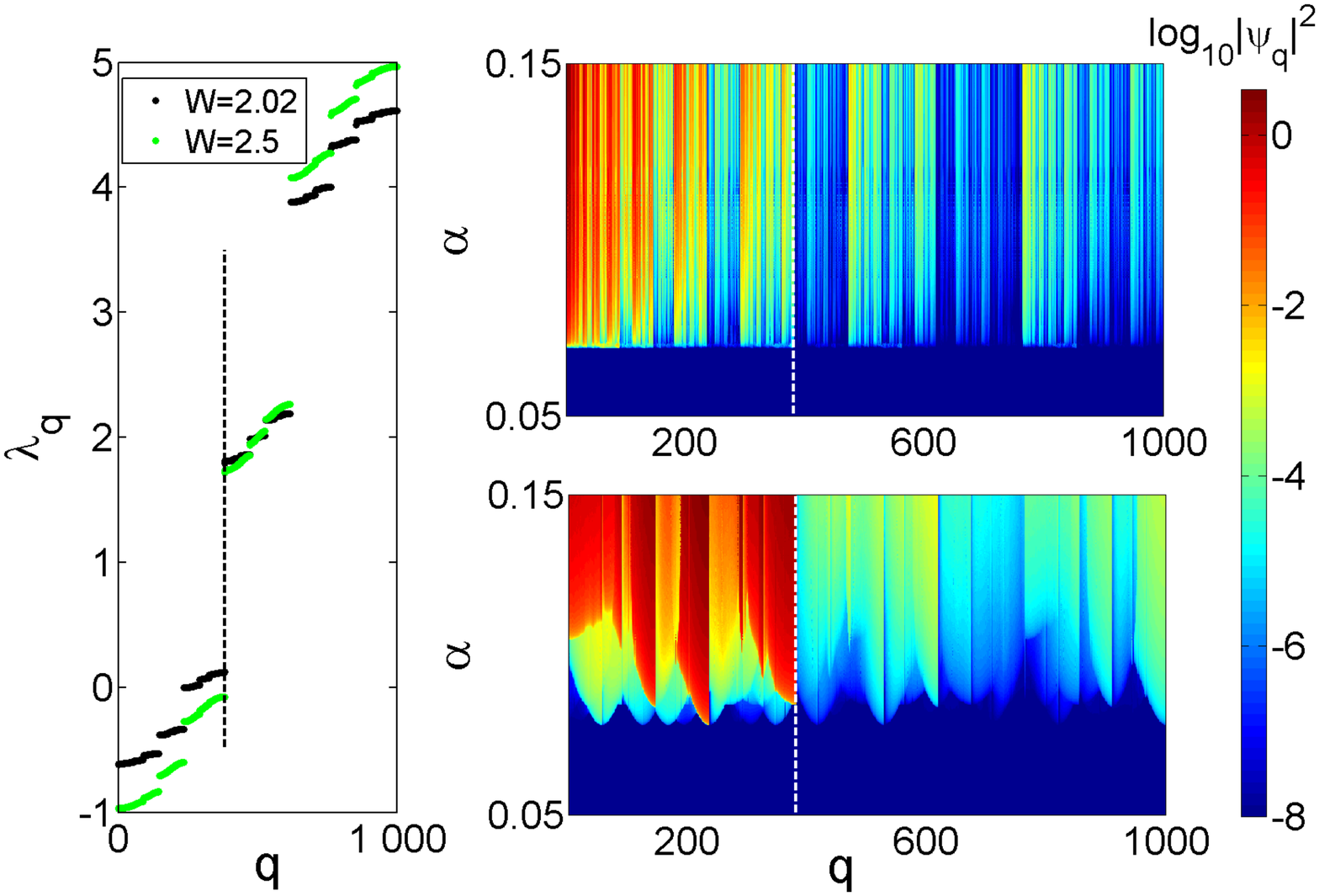}
\caption{(Color online) Linear mode space analysis, cf. Fig.\ref{fig:1}. Left panel: the linear spectrum in the insulating regime of the conservative Aubry-Andre limit, $\lambda_q$ in the increasing order, displaying three main bands for the given parameters. Right panels: color coded mode amplitudes, $\log_{10}|\psi_q|^2$ upon the increase of the pumping $\alpha$ for weak ($W=2.02$, top) and strong localization ($W=2.5$, bottom). Vertical dashed lines in the panels separate the lower band and indicate its dominating excitation.}
\label{fig:2}
\end{center}
\end{figure}

Figure~\ref{fig:1} presents characteristic results of numerical simulations for weak and strong localization, $W=2.02$ ($\xi\approx$100) and $W=2.5$ ($\xi\approx 4$). Profiles for different values of $\alpha$  were obtained as independent \textit{attractor} solutions, by setting system into an initial random low-energy state $|z_l(0)| \ll 1$ and letting it evolve until the corresponding  amplitude profile is stabilized \cite{attractor}. The key distinction of the attractor patterns from the disordered Anderson model \cite{we_arxiv} is the visually global excitation patterns in the weak localization regime (Fig.~\ref{fig:1}). At he same time, the structure of multiple localization peaks away from the metal-insulator transition looks similar to the Anderson system. 

Switching to the eigenmode basis we confirm the conjecture that the modes from the lower band get excited primarily and keep dominating with growing $\alpha$ (Fig.~\ref{fig:2}). We again notice that excitation of many modes occurs is grouped tightly in $\alpha$.  

\begin{figure}[t]
\begin{center}
\includegraphics[width=\columnwidth,keepaspectratio,clip]{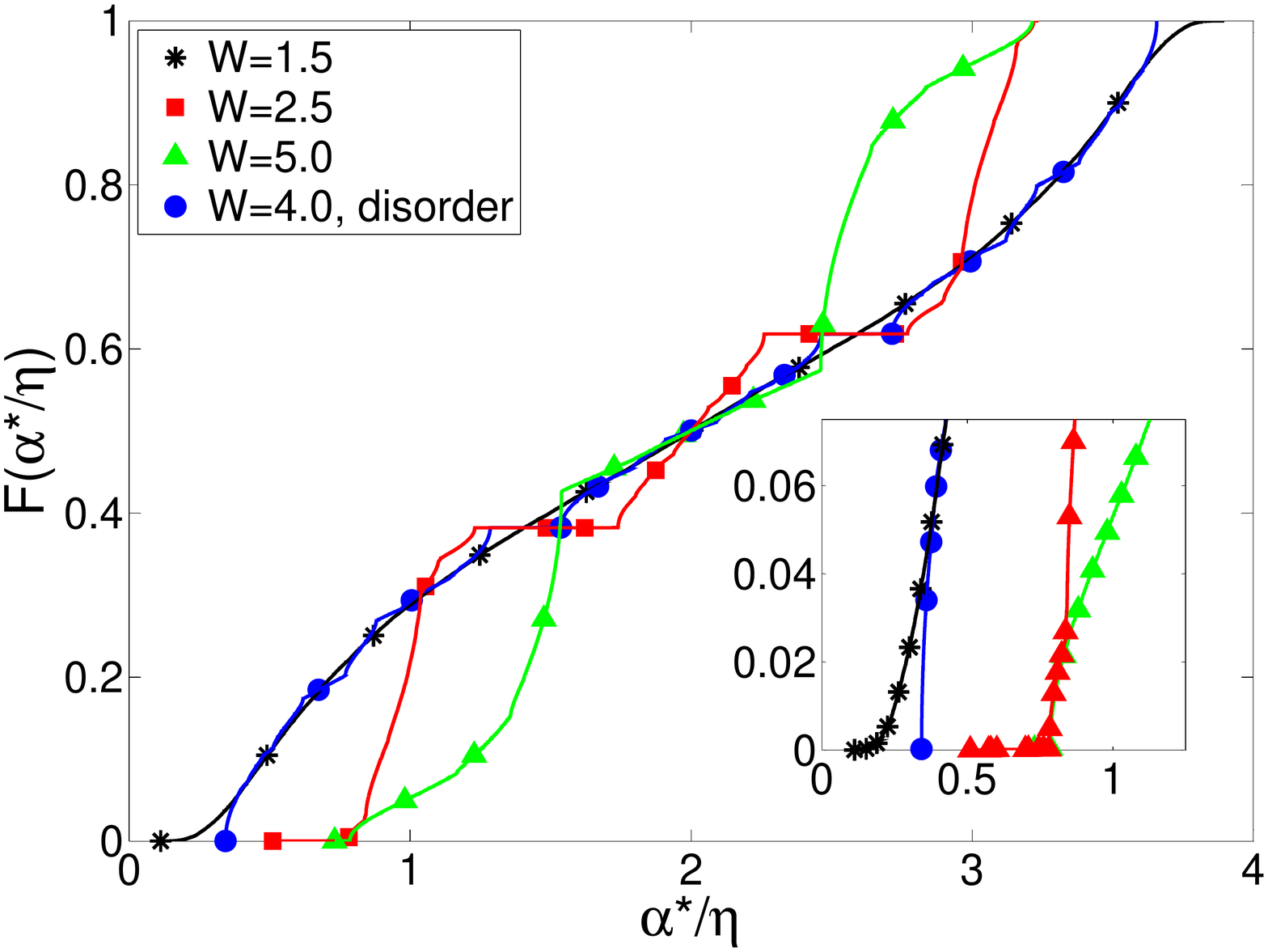}
\caption{(Color online) Integrated density distributions for individual mode excitation thresholds $F(\alpha^*/\eta)$ for incommensurate arrays with $W=1.5$ (extended linear states), $W=2.5$, and $W=5$ (localized states), and for the disordered array with $W=4$ (localized states). Inset: enlarged part of the main figure about excitation threshold, demonstrating different scaling. The system size is $N=5000$.}
\label{fig:4}
\end{center}
\end{figure}

To study the excitation transition in detail we first compute individual increments for the modes (\ref{eq:6}) and construct their integrated density distribution $F(\alpha/\eta)$. The results reveal the qualitative distinction between incommensurate and disordered lattices, that is the vertical slope of $F(\alpha/\eta)$ about the excitation point (Fig.~\ref{fig:4}). Such difference is inherited from the density of states for both models, the Anderson disorder producing Lifshitz tails, absent in the Aubry-Andre case. 

\begin{figure}[t]
\begin{center}
\includegraphics[width=\columnwidth,keepaspectratio,clip]{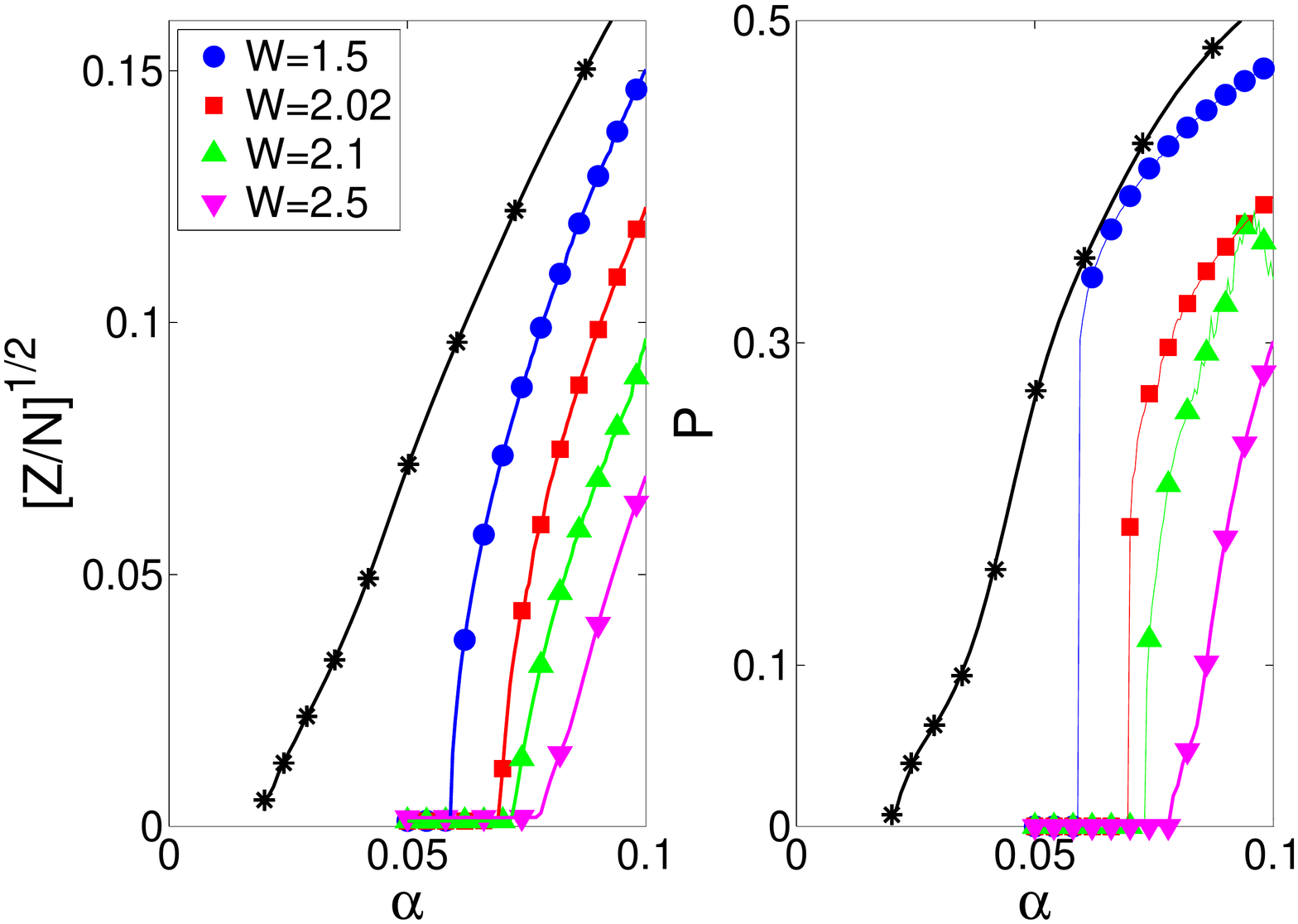}
\caption{(Color online) The average amplitude density $\sqrt{Z/N}$ (left panel) and participation number $P$ (right panel) in dependence on the pumping strength $\alpha$ for incommensurate lattices $W=1.5$ (extended linear states), $W=2.02$, $W=2.1$, $W=2.5$ (localized states), and the disordered array with $W=4$. The parameters are $\eta=0.1, N=500$.}
\label{fig:5}
\end{center}
\end{figure}

Going beyond single mode approximation we perform numerical integration of (\ref{eq:1a}) and calculate the time-averaged amplitude density $\sqrt{Z/N}$ and participation number (a quantity commonly used to estimate the number of effectively excited sites) normalized by the system size
\begin{equation}
\label{eq:16}
P=\left(\frac{1}{N}\sum |z_l|^4/Z^2\right)^{-1}
\end{equation}
as the functions of the pumping strength $\alpha$ (Fig. \ref{fig:5}).
Notably, passing from extended to localized linear states (at $W=2$), one observes that the slope of the $\sqrt{Z/N}$ as the function of $\alpha$ at the excitation threshold point decrease from about $1/2$, as for the extended states to about $1$, as realized for Anderson disorder. Moreover, in both metallic and insulating regimes the participation numbers show a diverging slope at this point $dP/d\alpha\rightarrow\infty$, featuring the second order phase transition, in contrast to the disordered case. In other words, incommensurate lattices in the weak insulating regime ($W<2.5$) present emerging oscillating patterns that are global in the array, manifesting simultaneous delocalization transition.

\begin{figure}[t]
\begin{center}
\includegraphics[width=\columnwidth,keepaspectratio,clip]{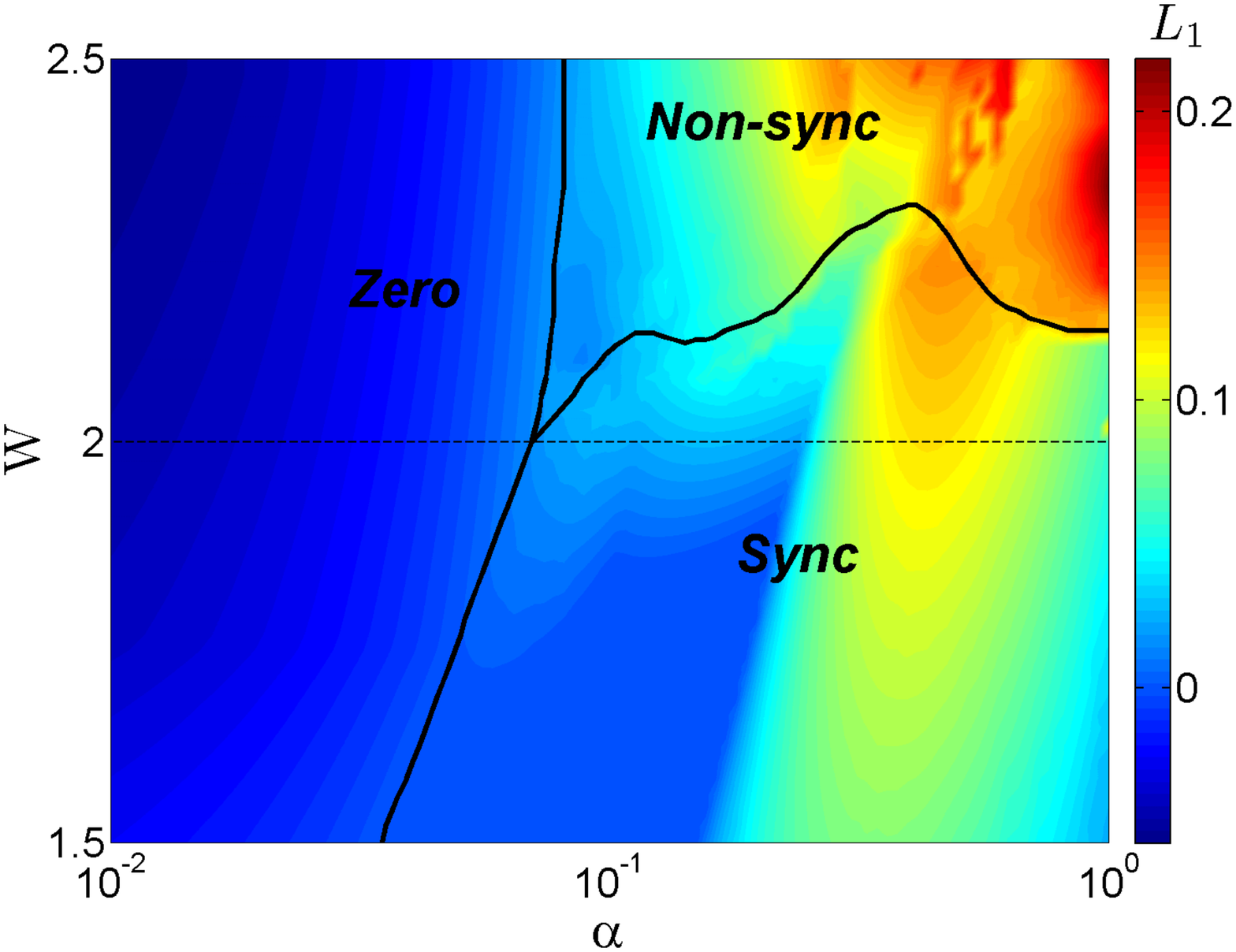}
\caption{(Color online) Pumping -- modulation depth parameter diagram: color coded largest Lyapunov exponent $L1$ depicts regular and chaotic regimes, solid lines separate globally stable zero, synchronized and non-synchronized dynamics (``Zero'', ``Sync'', and ``Non-sync'' labels), horizontal dashed line at $W=2$ indicates metal-insulator transition in the Aubry-Andre limit (\ref{eq:1c}). The parameters are $\eta=0.1, N=500$.}
\label{fig:6}
\end{center}
\end{figure}

Finally we characterize dynamical regimes of developed oscillations, calculating the largest Lyapunov exponent $L_1$
as a function of the pumping strength and modulation depth, and average individual oscillation frequencies, marking the onset of synchronization \cite{Pik_book} when the maximal cluster size exceeds ten percent of the array length, $N=500$, Fig. \ref{fig:6}.
The resulting diagram demonstrates the gradual increase of excitation threshold $\alpha^*=\alpha^*(W)$, when Aubry-Andre states are extended ($W<2$), and only weak variations above $W=2$ with $\alpha^*/\eta\approx1$, corroborating the above estimate for $W\gg2$. Chaotic region has a complex structure with non-monotonous dependence of $L_1(\alpha)$, and even temporary regularization for $1.7<W<1.9$ as $\alpha$ is increased. Instructively, the multi-mode dynamics that develops for $W>2$ once $\alpha>\alpha^*$ leads to a rapid transition to chaos. Synchronization has proved to be possible in both regular and chaotic regimes, lost as the modulation depth $W$ increases.

In conclusion, localization in active incommensurate systems exhibits qualitatively different features, as compared to disordered arrays. 
In the conservative limit they display transition from extended (metallic) to localized (insulating) states as the modulation depth is increased. We have found that the interplay of dissipation and pumping produces excitation of oscillations at finite pumping strength, corresponding to the linear modes from the lower frequency band. Remarkably, it manifests the features of the second order phase transition and the onset global oscillations not only from the metallic, but also from the weakly insulating conservative states. The Anderson attractors regime for disordered systems with sparse patterns of excited Anderson modes is reproduced in the strong localization limit only. Chaotic synchronized and non-synchronized oscillations develop as pumping is further increased.  

The spectrum of candidate systems, where the reported effects can be studied experimentally, includes lattices of polariton condensates \cite{P_reviews,berloff}, cavity-QED arrays with the cavities  filled up with two-level atoms or qubits \cite{qed}, systems of coupled Josephson junction \cite{Basko} or optical waveguide arrays \cite{Flach}.

T.L, G.O. and M.I. acknowledge support of the Russian Science Foundation No. 14-12-00811.


\begin{thebibliography}{99}
\bibitem{attractor} We do not exclude the possibility of multiattractor regimes in principle. However, 
 we observed single-attractor regimes in all performed numerical tests.
\end{thebibliography}
\end{document}